\documentclass[twocolumn,showpacs]{revtex4}
\usepackage{amssymb,epsfig,amscd}
\usepackage{graphicx}
\usepackage{dcolumn}
\usepackage{bm}
 
\usepackage{epsfig}

\def\be{\begin{equation}} 
\def\ee{\end{equation}}
\def\bq{\begin{eqnarray}} 
\def\eq{\end{eqnarray}}

\begin{document}

\title{Gapless color-flavor locked phase in quark and hybrid stars}

\author{A. Lavagno$^{a,b}$, G. Pagliara$^{a,c}$}

\affiliation{$^a$Dipartimento di Fisica, Politecnico di Torino, 10129 Torino, Italy}
\affiliation{$^b$INFN Sezione di Torino, 10125 Torino, Italy \\
$^c$INFN Sezione di Ferrara, 44100 Ferrara, Italy}

\begin{abstract} 

We study the effects of the gapless color-flavor locked (gCFL) phase on
the equation of state of strongly interacting matter in the range of
baryonic chemical potential involved in a compact star.  We analyze
the possibility of a phase transition from hadronic matter to gCFL
quark matter and we discuss, for different values of the strange quark
mass and diquark coupling strength, the existence of a gCFL phase in
quark or hybrid stars. The mass-radius relation and the structure of
compact stars containing the gCFL phase are shown and the
physical relevance of this superconducting phase inside a stellar
object is also discussed.

\end{abstract}

\pacs{26.60.+c, 25.75.Nq, 24.85.+p, 97.60.Jd}

\noindent

\maketitle

\section{Introduction}

Recent studies on the QCD phase diagram at finite densities and
temperatures have revealed the existence of a rich structure of the
phase diagram with several possible types of color superconducting
phases \cite{Rajagopal:2000wf,Alford:2001dt}.  These results are very
interesting in the study of the structure and formation of compact stellar
objects in which the central density may reach values up to ten times
the nuclear matter saturation density and therefore deconfinement of
quarks may take place.  Similar conditions may also be reached in
future heavy ion colliders, as at GSI, where it will be possible to
study the transition from hadronic matter to quark gluon plasma, and 
the possible low-temperature transition from hadronic matter to 
color superconducting quark matter.

The structure of the QCD phase diagram at high densities and
vanishing temperature (the conditions in a compact star)
depends strongly on the value of the strange quark mass $m_s$.  For
the two extreme cases, vanishing $m_s$ and very large values of $m_s$,
it is widely accepted that the three-flavor color-flavor locked (CFL)
phase and the two-flavor color superconducting phase (2SC) are,
respectively, the most favored phases
\cite{Rajagopal:2000wf,Alford:2001dt}. At intermediate values of
$m_s$, it is in general difficult to involve strange quarks in BCS
pairing due to their Fermi momentum, which is lower than that of up and
down quarks, therefore the CFL phase can form only if the CFL
superconducting gap, $\Delta_{CFL}$, is large enough
\cite{Rajagopal:2000ff}.  Recently, it has been shown that the CFL
phase can form only if the ratio $m_s^2/\mu \lesssim 2 \Delta_{CFL}$
\cite{Alford:2003fq}. At larger values of $m_s^2/\mu$, but not too
large values of $m_s$, the most energetically favored phase is the
so-called gapless CFL (gCFL) phase instead of the 2SC phase or
unpaired quark matter (UQM)
\cite{Alford:2004hz,Fukushima:2004zq,Ruster:2004eg}.  The gCFL phase
has the same symmetries as the CFL phase but there are two gapless
quark modes and a nonzero electron density. The existence of gapless
degrees of freedom makes the gCFL phase a conductor, at variance from
the CFL phase, and it is expected to have very different transport
properties with respect to the CFL phase.  The relevance of the gCFL phase
in the equation of state of strongly interacting matter at low
temperature is actually controversial and depends on the value of
the diquark coupling, on the strange quark mass and on the baryonic
chemical potential. In particular, it has very recently been shown
that on taking into account dynamical chiral symmetry breaking
within the NJL model, the 2SC phase would be the favored phase at low
densities if large values of the diquark coupling were used
\cite{Abuki:2004zk,Ruster:2005jc,Blaschke:2005uj}.  Although there are
uncertainties in the region of the QCD phase diagram in which the
gCFL phase occurs, as outlined in
Refs.\cite{Alford:2003fq,Alford:2004hz,Fukushima:2004zq}, the gCFL phase
may have a relevant role in compact stars because of the wide range of baryonic
chemical potential involved in these stellar objects.  A first study
of the effect of the presence of the gCFL phase in a compact star was
presented in Ref.\cite{Alford:2004zr}. In that paper the authors
compute the specific heat and neutrino emissivity of the gCFL
phase claiming that if the gCFL phase forms in an old compact star, it
should deeply affect its cooling. These results are obtained assuming
that the star is composed of a nuclear matter crust and a gCFL core at a 
fixed value of the quark chemical potential ($\mu=500$ MeV),
corresponding approximately to ten times the nuclear baryon
density, the density reached in the core of a compact star.  Actually,
a detailed investigation on the relevance of the gCFL phase on the
structure and formation of a compact star is still lacking in
literature.

The main goal of this paper is to study the possible existence of the
gCFL phase in the large range of the baryonic chemical potential involved
in compact stars. We will compute, for different values of the
strange quark mass and diquark coupling strength, the equation of
state (EOS) of the gCFL phase considering also the possibility of a
phase transition from hadronic matter to gCFL quark matter. The
resulting EOSs will then be used to obtain the structure of both quark
and hybrid stars.  This paper is organized as follows: in Sections II and
III, we study the relativistic hadronic EOS and the three quark
flavors EOS (gCFL-CFL), respectively;  in Sec. IV we discuss the
possibility of phase transitions from hadronic matter to quark
matter; in Sec.V we compute the mass-radius relations of quark and
hybrid stars; the conclusions are reported in Sec.VI.

\section{Equation of state of hadronic matter}

Concerning the hadronic phase, we use a relativistic self-consistent  
theory of nuclear matter in which nucleons interact through the nuclear force 
mediated by the exchange of virtual isoscalar and isovector mesons ($\sigma,\omega,\rho$)
\cite{Glendenning:1991es}.  
At $T=0$, in the mean field approximation, 
the thermodynamic potential $\Omega$ per unit volume can be written as
\begin{eqnarray}
\Omega=&-&\frac{1}{3 \pi^2}\sum_{B} \int_0^{k_{FB}}\!\!\!\! {\rm d}k
\frac{k^4}{E_{B}^\star(k)}+\frac{1}{2} m_\sigma^2\sigma^2 \nonumber \\
&+& \frac{1}{3} a \sigma^3+\frac{1}{4}b\sigma^4
-\frac{1}{2}m_\omega^2\omega^2-\frac{1}{2}m_{\rho}^2 \rho^2 \, ,
\end{eqnarray}
where the $\sum_{B}$ runs over the eight baryon species, ${E_{B}^\star(k)}=\sqrt{k^2+{M_{B}^\star}^2}$ and 
the baryon effective masses are ${M_{B}^\star}=M_{B}-g_\sigma \sigma$. 
The effective chemical potentials $\nu_B$ are given 
in terms of the thermodynamic chemical potentials $\mu_B$ and of the vector meson fields 
as follows
\be
{\nu_B}=\mu_B  - g_\omega\omega -t_{3B} g_{\rho}\rho \,  , 
\ee
where $t_{3B}$ is the isospin 3-component for 
baryon $B$ and the relation to the Fermi momentum $k_{FB}$ is provided by 
$\nu_B=\sqrt{k^2_{FB}+{M_B^\star}^2}$. 
The isoscalar and isovector meson fields 
($\sigma$, $\omega$ and $\rho$) are obtained as a solution of the field equations 
in mean field approximation and the related couplings ($g_\sigma$, $g_\omega$ and 
$g_\rho$) are the parameters of the model \cite{Glendenning:1991es,Glend:01kn,Knorren:1995ds}.

\begin{figure}[t]
\begin{center}
\includegraphics[scale=0.5]{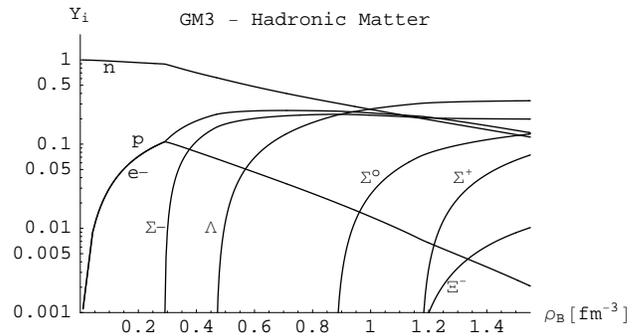}
\end{center}
\parbox{8cm}{
\caption{Particle fractions $Y_i$ of neutral and $\beta$-stable hadronic matter 
as a function of baryonic density $\rho_B$ for the GM3 
hadronic equation of state \cite{Glendenning:1991es}}.
\label{fighadronic}}
\end{figure}

In Fig. \ref{fighadronic} we display the relative concentrations of the various particle species 
$Y_i=\rho_i/\rho_B$ as a function of baryonic density $\rho_B$ by imposing charge neutrality 
and $\beta$-equilibrium for the GM3 parameter set \cite{Glendenning:1991es}. 

\section{Equation of state of the gapless CFL phase}

To compute the EOS of the gCFL phase we adopt the NJL-like formalism
of Refs.\cite{Alford:2003fq,Alford:2004hz,Fukushima:2004zq} in which the thermodynamic
potential per unit volume can be written as
\bq
\Omega&=& -\frac{1}{4 \pi^2} \int \mathrm{d}\,p\,\, p^2 \sum_j|\epsilon_j(p)|\rho_j(p)\nonumber\\
       &+&  \frac{1}{G}(\Delta_1^2+\Delta_2^2+\Delta_3^2)-\frac{\mu_e^4}{12 \pi^2} \, ,
\label{omega}
\eq
where $\Delta_1$, $\Delta_2$, $\Delta_3$ are the superconducting gaps
characterizing the gCFL phase (which reduce to a single gap in the CFL
phase), $G$ is the strength of the diquark coupling, $\epsilon_j(p)$
are the dispersion relations of quarks as in Ref.~\cite{Alford:2004hz} and $\mu_e$ is the 
electron chemical potential. Following the approximations used in
Refs.\cite{Alford:2003fq,Alford:2004hz}, the effect of $m_s$ is
introduced as a shift $-m_s^2/2\mu$ in the chemical potential for the strange quarks
and the contributions of antiparticles is neglected. The first approximation 
is applicable for values of $m_s$ smaller than the chemical potential and therefore  
we will use in this paper typical values of $m_s$ of $150-200$ MeV.   
Concerning the antiparticles, as already remarked in Ref.~\cite{Alford:2004hz}, 
neglecting their contribution can lead to incorrect values of the thermodynamic variables.
In that paper, this problem does not play a relevant role because the free energy 
differences relative to  UQM are presented.
Here we are going to investigate the structure of a compact star and we need 
to compute the variation of the thermodynamic potential as a function of the chemical potential. 
To this end, following \cite{fetter,Alford:1998mk}, we have introduced in Eq.(\ref{omega})
the quasiparticle probabilities 
\be
\rho_j(p)=\frac{1}{2}\left(1-\frac{\tilde{\epsilon}_j(p)}{\epsilon_j(p)}\right) \, ,
\ee 
where $\tilde{\epsilon}_j(p)$ are the dispersion relations with vanishing
gaps. To assure the convergence of the integral in Eq.(\ref{omega}), a form factor 
$f = (\Lambda^2/(p^2+\Lambda^2))^2$, which multiplies the gaps, 
is introduced in the dispersion relations $\epsilon_j(p)$.
The form factor was fixed to mimic the effects of the asymptotic freedom 
of QCD \cite{Alford:1998mk} and the parameter $\Lambda$ was fixed at
a value of $800$ MeV. For the CFL phase, this procedure leads to results in agreement 
with previous calculations performed using the simplified model of 
Refs.~\cite{Alford:2001zr,Alford:2002rj}. 

To describe the matter of a compact star, the conditions of chemical equilibrium between quarks, 
charge neutrality and color neutrality must be imposed.
The chemical equilibrium conditions (which also include $\beta$-stability) 
allow the expression of each quark chemical potential
$\mu_{c f}$ ($c$ and $f$ are the indexes of color and flavor, respectively) 
as functions of quark (baryonic) chemical potential $\mu$, electron
chemical potential $\mu_e$ and the two chemical potentials, $\mu_3$
and $\mu_8$, associated to the $U(1)\times U (1)$ subgroup of the
color gauge group (see Ref.~\cite{Alford:2004hz} for details). The
color and electric charge neutrality are imposed by the following three equations:
\be
\frac{\partial \Omega}{\mu_3} = 0, \hspace{0.5cm} \frac{\partial \Omega}{\mu_8} = 0, \hspace{0.5cm} 
\frac{\partial \Omega}{\mu_e} = 0 \, .
\ee
Moreover, the thermodynamic potential must be minimized with respect to the gap
parameters and therefore we have to impose the three additional conditions 
\be
\frac{\partial \Omega}{\Delta_1} = 0,\hspace{0.5cm} \frac{\partial \Omega}{\Delta_2} = 0, \hspace{0.5cm}
\frac{\partial \Omega}{\Delta_3} = 0 \, .
\label{d3}
\ee
The above equations allow us to compute the thermodynamic potential and all the thermodynamic 
variables as a function of the quark
chemical potential only. On solving these equations, we obtained 
results in agreement with the ones  shown in Ref.\cite{Alford:2004hz} concerning
the dispersion relations, the gaps and the chemical potentials as a function of $m_s^2/\mu$.

\begin{figure}[t]
\begin{center}
\includegraphics[scale=0.5]{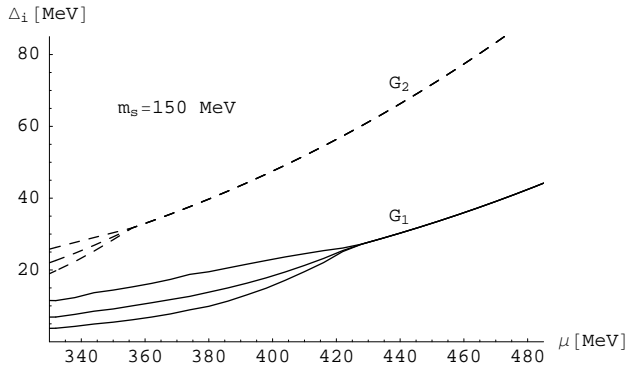}
\end{center}
\parbox{8cm}{
\caption{Gap parameters as a function of the
quark chemical potential for two different values of  
diquark coupling $G_1$ and $G_2$ and for a fixed value of strange quark mass
$m_s=150$ MeV. The larger the value of $G$, the larger the window
of the chemical potential in which CFL phase occurs.
 }
\label{gap}}
\end{figure}

\begin{figure}[t]
\begin{center}
\includegraphics[scale=0.5]{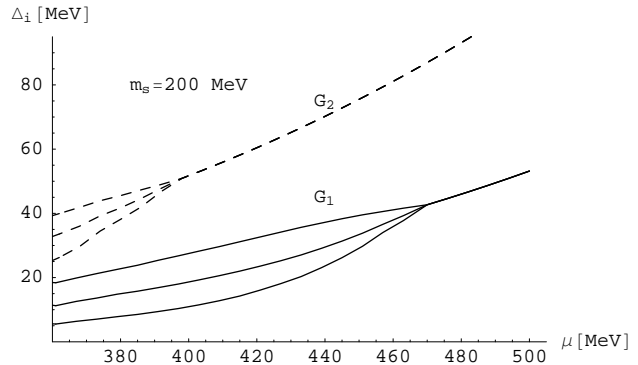}
\end{center}
\parbox{8cm}{
\caption{The same as Fig.\ref{gap}, with $m_s=200$ MeV. }
\label{gap1}}
\end{figure}

In Figs.~\ref{gap} and \ref{gap1} the gap parameters are displayed as
functions of quark chemical potential $\mu$ for two different values of $m_s$. 
Here we have used two values of
the diquark coupling $G_1$ and $G_2$ corresponding, respectively, to values of
$\Delta_{CFL} \sim 40$ and $\Delta_{CFL} \sim 100$ MeV at $\mu = 500$
MeV and $m_s = 150$ MeV. It is interesting to observe
that the window in which the gCFL phase appears depends noticeably on the value of
the diquark coupling and, in particular, it increases with $m_s$ and
decreases with $G$ (see Figs.~\ref{gap} and \ref{gap1}). This confirms
the general argument for which the transition from gCFL to CFL occurs when
$m_s^2/\mu\simeq 2 \Delta_{CFL}$.

\begin{figure}[t]
\begin{center}
\includegraphics[scale=0.5]{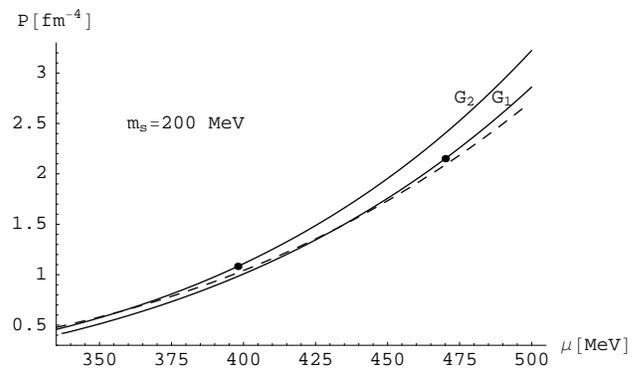}
\end{center}
\parbox{8cm}{
\caption{Pressure as a function of the quark chemical potential
for UQM (dashed line) and for the (g)CFL phase (solid
lines) for two values of diquark coupling $G_1$ and $G_2$. The dots 
represent the onset of the gCFL-CFL phase for the two coupling parameters $G_1$ and $G_2$.
The bag constant is fixed at a value $B^{1/4}=150$ MeV.
\label{pmu}}}
\end{figure}

\begin{figure}[t]
\begin{center}
\includegraphics[scale=0.5]{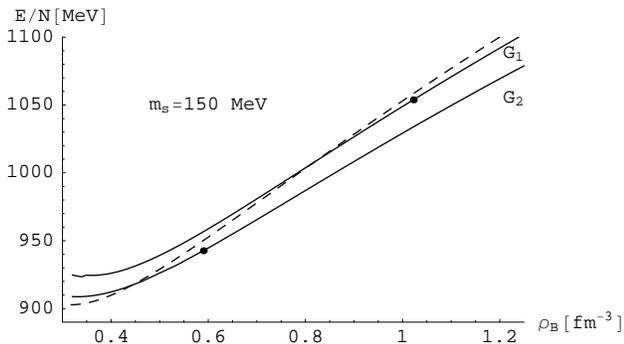}
\end{center}
\parbox{8cm}{
\caption{Energy per baryon displayed as a function of 
baryon density for fixed values of the bag constant $B^{1/4}=150$ MeV and the strange
quark mass. The dashed line corresponds to UQM and the
solid lines correspond to (g)CFL EOSs for the two choices of 
diquark coupling $G_1$ and $G_2$. The dots represent the onsets of the gCFL-CFL phase.  }
\label{eneperbar}}
\end{figure}

In Figs.~\ref{pmu} and \ref{eneperbar} we compare the EOSs of the (g)CFL and
UQM phases. A bag constant ($B^{1/4} = 150$ MeV in Figs.~\ref{pmu} and \ref{eneperbar})
has been added to the thermodynamic potential to 
simulate quark confinement, as in the MIT bag model.
In Fig.~\ref{pmu}, we display the pressure of the gCFL and UQM phases for two
choices of coupling $G$ at a fixed value of $m_s$.  Let us
observe that the smaller values of the diquark coupling imply a larger
chemical potential window in which gCFL occurs; however, only a
part of this window is favored in comparison with UQM.  In fact, from
Fig.~\ref{pmu} it can be observed that gCFL is more energetically
favored than UQM for a window of about $\Delta\mu \simeq 30-50$ MeV.  It is also to be noted  
that the CFL phase is always favored at large chemical potentials.  A
similar behavior was also found in Ref.~\cite{Ruster:2005jc} on taking
into account the dynamic generation of quark masses.  Calculation of
the energy per baryon as a function of the baryon density is shown in
Fig.~\ref{eneperbar}.  It is also evident in this figure that gCFL is
energetically favored with respect to UQM in a strict region of baryon
density.
The EOSs displayed in Fig. \ref{eneperbar} satisfy the Bodmer-Witten 
hypothesis \cite{Bodmer:1971we,Witten:1984rs,Drago:2001nq} which states that the true and absolute 
ground state of the strong interaction is quark matter. 
As we will see in Sec. V, this will lead to stellar objects
entirely occupied by quark matter. 


\section{Phase transition from hadronic matter to the gapless CFL phase}

The region of the QCD phase diagram in which the transition from
hadronic matter to quark matter should occur, is little known.  In
particular, it is not yet known if there is a direct transition from
hadronic matter to the ground state of QCD, i.e. the CFL phase,
or if there is an intermediate density window in which another quark
phase may appear. In several papers \cite{Alford:2002rj,Banik:2002kc,Baldo:2002ju,Drago:2004vu},  
it was assumed that this transition is direct by using 
an approximate equation of state valid for large values of
the gap parameters. However, as already remarked, the gCFL phase may be a valid candidate in
connecting hadronic matter to the CFL phase at intermediate densities and it is 
very important to see if the first order phase transition between hadronic
matter and gCFL possibly occurs via an intermediate density window
of mixed phase.  

The EOS appropriate to the description of a compact
star has to satisfy $\beta$-stability conditions. This implies the
existence of two conserved charges, the baryonic charge and the
electric charge. When the Gibbs conditions are applied in presence of
more than a single conserved charge, the technique developed by
Glendenning \cite{Glend:01kn} has to be adopted and the equivalence of
the baryon and the charge chemical potentials in the two phases 
must be imposed.  Moreover, the electric neutrality is required as a
global condition \footnote{Color neutrality is instead imposed as a local condition 
because color is clearly confined in the hadronic phase.}. The corresponding equation reads:
\be 
(1-\chi) \rho_c^H + \chi \rho_c^{gCFL}-\rho_e =0 \, ,
\label{neutral}
\ee 
where $\chi$ is the volume fraction of the quark phase, $\rho_c^H$ is
the charge density of the hadronic phase given by protons and charged
hyperons if present, $\rho_c^{gCFL}$ is given by the density of the
gapless $bu$ quarks (see Ref.~\cite{Alford:2004hz} for details),   
$\rho_e$ is the density of electrons.   

\begin{figure}[t]
\begin{center}
\includegraphics[scale=0.5]{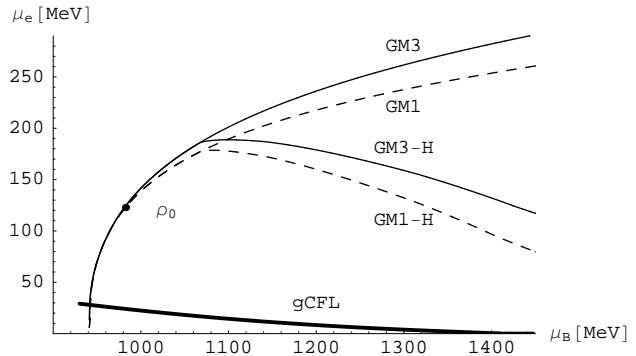}
\end{center}
\parbox{8cm}{
\caption{Electrons chemical potential as a function of the
baryon chemical potential for the pure neutral hadronic and gCFL
phases. Solid and dashed lines correspond to the nuclear (np) and hadronic (npH) 
matter in models GM1 and GM3 of Ref.\cite{Glendenning:1991es}. The
thick solid line corresponds to the gCFL phase with $m_s = 200$ MeV and for diquark coupling $G_1$.
The dot labeled with $\rho_0$ corresponds to nuclear matter saturation density.}
\label{chemical}}
\end{figure}

In computing the first critical density ($\chi =0$),
Eq.(\ref{neutral}) coincides with the charge neutrality equation
for the pure hadronic matter. 
Since the transition to a mixed phase of hadronic-quark matter
eventually takes place at densities larger than nuclear matter density $\rho_0$,  
the electron chemical potential, as can be seen in Fig.\ref{chemical}, 
is larger than $\mu_e \sim 100$ MeV. 
It turns out that for such a high value of $\mu_e$, 
the resulting splitting of quarks' Fermi momenta is too large for the given
diquark coupling strength to enable gCFL diquark pairing.  
Moreover, in the mixed phase UQM may contribute with a total
negative charge density which neutralizes the positive charge of
protons and the chemical potential of electrons 
decreases to few MeVs. Since the gCFL phase  
contributes with a positive charge density given by the gapless $bu$
quarks, the chemical potential of electrons must therefore have large
values to neutralize the positive charges of hadron and quark phases.
Let us remark that at this stage the formation of meson condensates \cite{Kaplan:2001qk}
has not been considered in the gCFL phase. The formation of Goldstone bosons may influence
noticeably the mixed phase at intermediate chemical potentials.     
Unfortunately, no investigations on this possibility are present in literature.  
Moreover, if the 2SC phase appears at low densities 
(i.e. for large values of $m_s$ and the diquark coupling parameters)
the creation of the mixed phase may be favored. 
The relevance of these possibilities lies beyond the scope of this paper and 
will be investigated in future works. 

\begin{figure}[t]
\begin{center}
\includegraphics[scale=0.5]{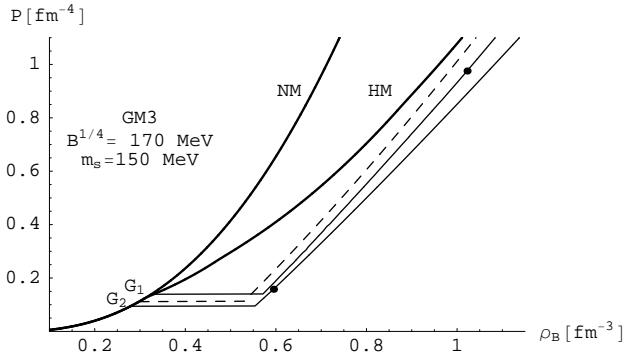}
\end{center}
\parbox{8cm}{
\caption{Pressure as a function of baryon density 
for different choices of parameters $G$.  The thick solid lines
correspond to the nuclear and hadronic matter GM3 EOSs (NM and
HM). The dashed line represents the case in which
a phase transition from hadronic matter to UQM is considered using the 
Maxwell construction.  The thin solid lines 
are related to the phase transition with the gCFL
phase for two choices of diquark pairing $G$. The dots on the
solid lines correspond to phase transition from gCFL to CFL phases
($G_1$ and $G_2$ have the same values as in Fig.~\ref{gap} and
\ref{gap1}).}
\label{max}}
\end{figure}

\begin{figure}[t]
\begin{center}
\includegraphics[scale=0.5]{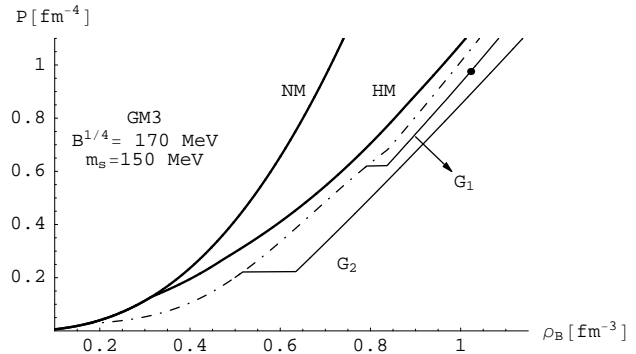}
\end{center}
\parbox{8cm}{
\caption{Pressure as a function of baryon density 
for the scenario in which a first phase transition from hadronic
matter to quark matter occurs via a mixed phase and then a second
phase transition (here computed using the Maxwell construction) occurs
from this mixed phase to the gCFL phase.  
The dot-dashed line
is related to the hadron-UQM mixed phase. The thin solid
lines represent the phase transitions from the mixed phase to the
gCFL phase for the two diquark couplings. 
Notice that in the case of $G_1$ a gCFL phase window is still present.}
\label{max2}}
\end{figure}

If Gibbs conditions cannot be realized, the phase transition between
hadronic matter and gCFL may occur with a discontinuity in the baryon
density by means of the Maxwell construction. This would be the case
even if a large surface tension between the hadronic and quark phases 
exists. In Fig.~\ref{max} the EOSs obtained using the Maxwell
construction are displayed. Concerning the effect of the diquark
coupling, the larger the value of $G$ the smaller the values of
the two critical densities. Notice that the density window in which
the gCFL phase appears (the dots on the solid curves indicate the onset of
the gCFL-CFL phase) decreases sharply with increases in the value of
$G$. The curve corresponding to the transition from hadronic matter to
UQM is shown in the same figure using the Maxwell construction (dashed
line). 
As we can see from Fig.~\ref{max}, 
if the transition from hadronic matter to quark matter occurs
with discontinuity in the baryon density, 
the EOS of the gCFL phase labeled with $G_2$ is favored with respect to
the UQM EOS.  If instead we take into account the possibility of the  
formation of a mixed phase, the transition from hadronic matter to quark matter
would occur via a hadron-UQM phase transition and then at
larger densities the gCFL and CFL phases may appear. This scenario is shown
in Fig.~\ref{max2} from which we can observe that for large diquark
coupling ($G_2$) the transition from the mixed phase to the (g)CFL
phase occurs above the onset of the gCFL-CFL phase and for small 
diquark coupling ($G_1$) a window of the gCFL phase is instead present. 
Concerning the hyperons, within this choice of parameters the
transition from hadronic matter to gCFL quark matter occurs before reaching the
threshold of the formation of hyperons. If we use larger values of the
bag constant the first critical density may be larger than the
threshold of the formation of hyperons.  In that case, however, the
phase transition would involve the CFL phase directly.

\section{Mass-radius relation of compact stars}

\begin{figure}[t]
\begin{center}
\includegraphics[scale=0.5]{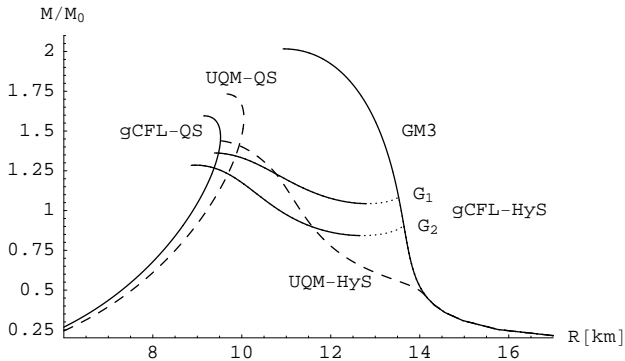}
\end{center}
\parbox{8cm}{
\caption{Mass radius relations of quark stars (QS), hybrid stars (HyS) and hadronic stars (GM3)  
are displayed for different EOSs. The initial part of the branch of gCFL hybrid stars
corresponds to unstable star configurations (dotted lines ). 
As the fraction of gCFL increases in the core
of hybrid stars the corresponding stars are stable (solid lines).
The EOSs used to compute the mass-radius relations are the same of those in 
Figs.~\ref{eneperbar} and \ref{max}.
In particular, concerning quark stars, the curve labeled gCFL-QS corresponds
to the gCFL EOS with diquark coupling $G_2$.
\label{massaraggio}}}
\end{figure}

The EOSs analyzed in the last sections for the gCFL phase can be used
to compute the structure of quark and hybrid stars. The EOS is an
input function needed to solve the Tolman-Volkoff-Oppenheimer system
of equations.  The mass-radius relations for quark stars and hybrid
stars are displayed in Fig.~\ref{massaraggio}. Concerning quark stars, 
the effect of the presence of the gCFL phase is not very appreciable and leads to a
small reduction of the radius of the star (see solid and dashed lines
in Fig.~\ref{massaraggio} labeled with gCFL-QS and UQM-QS) and it can
hardly be considered as a signature for the presence of the gCFL phase
in quark stars. 
Concerning hybrid stars, the way the transition from hadronic matter
to gCFL occurs, i.e. with a density discontinuity, is also reflected also in
the mass-radius relations of gCFL hybrid stars.  Stars in which a very
small fraction of the gCFL phase is present in the core are in fact
unstable, as can be seen from the initial part of the line
corresponding to the branch of hybrid stars (see dotted lines labeled
$G_1$ and $G_2$ in Fig.~\ref{massaraggio}).  On increasing the
fraction of quark matter in the volume of the stars, the radii decrease 
and stable configurations are obtained (see solid lines in
Fig.~\ref{massaraggio}).  As suggested in Ref.~\cite{third}, this
effect is explained by considering that the adiabatic index of the EOS
near the two critical densities is vanishing (using the Maxwell
construction) and therefore the pressure is too weak a function of
energy density to sustain stability. Even when using Gibbs conditions, 
similar instabilities are obtained within particular choices of the
parameters as shown in Ref.~\cite{Schertler:2000xq}.

\begin{figure}[t]
\begin{center}
\includegraphics[scale=0.5]{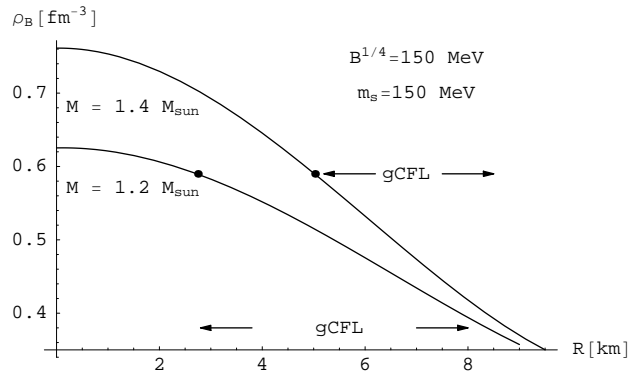}
\end{center}
\parbox{8cm}{
\caption{The baryon density profiles are shown 
for two quark stars having a mass of $1.4 M_\odot$ and $1.2 M_\odot$. 
The region of the stars indicated by the arrows
is composed of the gCFL phase. The diquark coupling used in the corresponding EOS is $G_2$. }
\label{struttureq}}
\end{figure}

\begin{figure}[t]
\begin{center}
\includegraphics[scale=0.5]{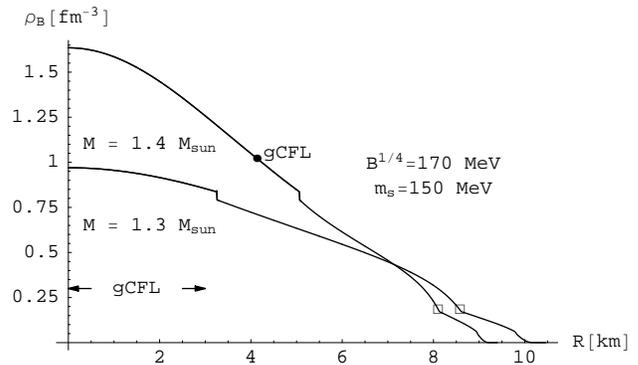}
\end{center}
\parbox{8cm}{
\caption{ The baryon density profiles are shown 
for two hybrid stars having a mass of $1.4 M_\odot$ and 
$1.3 M_\odot$. The dot indicates the onset of gCFL-CFL and the squares indicate
the beginning of the mixed phase. The diquark coupling used in the corresponding EOS is $G_1$.}
\label{strutturehyb}}
\end{figure}

It is also important to study the composition of quark and hybrid
stars in which the gCFL takes place. In Figs.~\ref{struttureq} and
\ref{strutturehyb}, the baryon density profile inside quark stars and
hybrid stars are shown, respectively. In quark stars, the baryon
density does not in general vary very much with the radius of the
star. Therefore, it is possible that almost all the star may actually
be composed of the gCFL phase (see Fig.~\ref{struttureq} for the $1.2
M_\odot$ star) with a small core of the CFL phase present.  For more
massive quark stars, the fraction of the volume occupied by the CFL
phase increases (see Fig.~\ref{struttureq} for the $1.4 M_\odot$
star). In both cases we expect the transport properties of the
matter of the star to be determined by the gCFL phase because the CFL phase is
essentially passive with all quarks gapped. In general, the presence
of a color superconducting phase has relevant effects on the cooling
of the stars as shown in Refs.~\cite{Alford:2004zr,Shovkovy:2002kv,Grigorian:2004jq}.
Moreover, the presence of the gCFL phase can also be very important in
the study of r-mode instability which imposes rather severe limits on the
highest rotation frequency of pulsars \cite{Friedman:1997uh,Andersson:2000mf}. 
Indeed, we expect the gapless modes of the gCFL phase to play 
an important role in the bulk and shear viscosity of the star \cite{Drago:2003wg}. 
For instance, in Ref.~\cite{Madsen:1999ci} it is shown that the existence of pure
CFL quark stars is ruled out by the existing data on pulsars because
of r-mode instability \footnote{Notice that in Ref.~\cite{Manuel:2004iv} the computation of the
contribution to the shear viscosity of phonon-phonon scattering leads to larger values 
of the viscosity and therefore this could help to stabilize the star.}. 
A crust of the gCFL phase may help to stabilize
the star and therefore quark stars may again be considered as
possible stellar objects. Concerning hybrid stars, their structures
are shown in Fig.~\ref{strutturehyb} using the EOS of Fig.~\ref{max2}
for diquark coupling $G_1$. As in the case of quark stars, the
volume occupied by the gCFL phase varies with the mass of the star.
In this case a crust of hadronic matter and a large layer of mixed
phase are present. Depending on the mass of the star, a narrow layer of
the gCFL phase ($M = 1.4 M_\odot$) or a core of the gCFL phase may form ($M =
1.3 M_\odot$). As suggested in Ref.\cite{Alford:2004zr}, the core or
the layer of the gCFL phase, which has a high heat capacity, can keep the
star warm for a long time, thus affecting the cooling process of the star.

\section{Conclusions}

In this paper we studied the relevance of the presence of the gCFL
phase in compact stars. We first computed the equation of state
of the gCFL phase and compared it with the unpaired quark matter equation
of state. We then investigated the phase transition from hadronic
matter to the gCFL phase. We found that the possibility of the formation of a
mixed phase between hadronic matter and the gCFL phase is 
hindered by the existence in the hadronic phase of a large number of electrons
which destroy gCFL pairing in the mixed phase. Therefore, the Maxwell
construction was adopted to connect the hadronic to the gCFL phase
and we found that the gCFL phase is energetically favored only in a
narrow region of the chemical potential. Finally, we studied the effect of
the presence of the gCFL phase in both quark and hybrid stars. Although the gCFL phase 
does not sensibly modify the global properties of stars, masses
and radii, we have shown that such a superconducting phase may occupy
a wide region inside the star and therefore may play a crucial
role in the calculation of the transport properties of the matter of
the star with very important physical implications. These results may 
stimulate more detailed studies on the effects of the gCFL phase in
the cooling and in the determination of the bulk and shear viscosity
inside a compact star, thus leading to a better understanding of the mechanism
acting in the suppression of r-mode instabilities in compact
stars.  Another interesting phenomenological signature of the presence
of the gCFL phase in compact stars may be provided by analyzing the role of
this phase in explosive phenomena such as Supernovae and gamma-ray
bursts. It is in fact well known that first order phase transitions in
compact stars can release huge amounts of energy \cite{Berezhiani:2002ks}. It may  
therefore be very important to see if the first order phase
transition from unpaired quark matter to the gCFL phase can occur 
during the life of a compact star. A hypothetical phase
transition from unpaired quark matter to gCFL quark matter may help
to explain the complex time structure of some gamma ray bursts
\cite{Lazzati:2005nc} in an astrophysical scenario in which the
central engine of gamma ray bursts is given by conversions between
different phases of strongly interacting matter.

\section{Acknowledgments}
It is a pleasure to thank A. Drago and P. Quarati for fruitful discussions and comments.

\bibliography{references}
\bibliographystyle{apsrev}

\end{document}